\documentstyle[12pt]{article}
\def\beq{\begin{equation}}
\def\eeq{\end{equation}}

\def\beqn{\begin{eqnarray}}
\def\eeqn{\end{eqnarray}}
\relax
\def\sp#1{{\rm Tr}\biggl( #1 \biggr)}

\def\ra{\rightarrow}
\def\prd#1#2#3{{\it Phys. Rev.} {\bf D#1} #2 (19#3)}
\def\pl#1#2#3{{\it Phys. Lett.} {\bf #1B} #2 (19#3)}
\def\np#1#2#3{{\it Nucl. Phys.} {\bf B#1} #2 (19#3)}
\def\prl#1#2#3{{\it Phys. Rev. Lett.} {\bf #1} #2 (19#3)}

\begin{document}
\begin{titlepage}
\def\ba{\begin{array}}
\def\ea{\end{array}}
\def\thefootnote{\fnsymbol{footnote}}
\begin{flushright}
	FERMILAB--PUB--93/218--T\\
	July, 1993
\end{flushright}
\vfill
\begin{center}
{\large \bf SIGNALS FOR PARITY VIOLATION IN THE\\
ELECTROWEAK SYMMETRY BREAKING SECTOR}\\
\vfill
	{\bf S.~Dawson$^{(a)}$}\footnote{
Work supported by the U.S. DOE under contract DE-AC02-76CH00016.}
        {\bf  and G.~Valencia$^{(b,c)}$}\\
{\it  $^{(a)}$ Physics Department,
               Brookhaven National Laboratory,  Upton, NY 11973}\\
{\it  $^{(b)}$ Theoretical Physics,
               Fermi National Accelerator Laboratory,
               Batavia, IL 60510}\\
{\it  $^{(c)}$ Department of Physics,
               Iowa State University,
               Ames, IA 50011}\\
\vfill
\end{center}
\begin{abstract}

We consider the possibility of observing a parity violating but $CP$
conserving interaction in the symmetry breaking sector of  the
electroweak theory. We find that the best probe for such an interaction
is a forward-backward asymmetry in $W^+W^-$ production from polarized
$e^-_R e^+_L$ collisions. An observable asymmetry would be strong
evidence against a custodial $SU(2)$ symmetry. We also discuss the
effects of such an interaction in future $e^- \gamma$
colliders as well as in rare decays of $K$ and $B$ mesons.

\end{abstract}

\end{titlepage}

\clearpage

\section{Introduction}

The standard model of electroweak interactions has now
been tested thoroughly in a number of experiments. The
only missing ingredients are the top-quark and the Higgs-boson.
Whereas we expect that the top-quark will be found in the
near future, the same cannot be said for the Higgs-boson.
The Higgs-boson in the standard model is responsible for
the breaking of electroweak symmetry, and experiments
conducted thus far have not tested directly the energy
scales at which the symmetry breaking is thought to occur.

There are many different physics possibilities that could be
responsible for the breaking of the  electroweak symmetry. This
makes it interesting to parameterize the symmetry breaking
sector of the theory in a model independent way, and to explore
the sensitivity of present and future experiments
to the new physics. In general, one can divide the possibilities
for the new physics into two classes. It is possible for the new
interactions to remain weakly coupled. Such models typically contain
new particles in the few-hundred GeV mass range.
Examples are models with low energy supersymmetry \cite{hhg}.
It is also possible that there are no new particles below a
few TeV and that the electroweak interactions become strong.
We will focus on the second possibility, although some of our
results apply in the first case as well.

We start from the minimal standard model without a Higgs
boson. This model can be written as the usual standard model,
but replacing the scalar sector with the effective Lagrangian
\cite{longo}:
\beq
{\cal L}^{(2)}={v^2 \over 4}\sp{D^\mu \Sigma^\dagger D_\mu \Sigma}.
\label{lagt}
\eeq
The matrix $\Sigma \equiv \exp(i\vec{w}\cdot \vec{\tau} /v)$, contains the
would-be Goldstone bosons $w_i$ that give the $W$ and $Z$ their
mass via the Higgs mechanism. They interact with the $SU(2)_L
\times U(1)_Y$ gauge bosons in a way dictated by the covariant
derivative:
\beq
D_\mu \Sigma = \partial_\mu \Sigma +{i \over 2}g W_\mu^i \tau^i
-{i \over 2}g^\prime B_\mu \Sigma \tau_3.
\label{covd}
\eeq
Eq.~\ref{lagt} is thus an $SU(2)_L \times U(1)_Y$ gauge
invariant mass term for the $W$ and $Z$. The physical masses
are obtained with $v \approx 246$~GeV. This non-linear realization
of the symmetry breaking sector contains the same low energy physics as
the minimal standard model when the Higgs-boson is taken to be very
heavy \cite{longo}. It is a non-renormalizable interaction
that is interpreted as an effective field theory, valid below
some scale $\Lambda$.
The details of the physics that break electroweak symmetry determine
the next-to-leading order effective Lagrangian. At energies small
compared to $\Lambda$, it is sufficient to consider those terms
that are suppressed by $E^2/\Lambda^2$ with respect to Eq.~\ref{lagt}.

We have previously discussed the case in which the new physics
contains a custodial $SU(2)$ global symmetry that is broken only
by the hypercharge coupling $g^\prime$ and by the mass splittings
in the left handed $SU(2)$ fermion doublets \cite{bdv}. Furthermore, we
specialized to the case of very high energy experiments in which
the scalar interactions are stronger than the gauge interactions
and it is consistent to set all the custodial $SU(2)$ violating
counterterms in the next-to-leading order effective Lagrangian to zero.

We now want to extend that analysis and study the effects of
custodial $SU(2)$ breaking counterterms. The one with the minimum
number of derivatives, two, is:
\beq
{\cal L}^{(2)}={1 \over 8} \Delta \rho v^2 \biggl[
\sp{\tau_3 \Sigma^\dagger D_\mu \Sigma}\biggr]^2.
\eeq
This term describes deviations of the $\rho$ parameter from
one\footnote{Experimentally,
$\Delta \rho =0.0016\pm 0.0032 $ \cite{rhop}.}  and has been
studied at length in the literature.
Unfortunately, there are many operators with four derivatives that break
the custodial symmetry
in the next-to-leading effective Lagrangian, making a general study
quite complicated. A complete set of these operators has been given in
Ref.~\cite{longo,appel}.  For specific problems, however, one finds that
only a few operators are relevant. For example, for physics at LEP,
only one of them contributes at tree-level to the
so called ``oblique'' electroweak corrections expected to dominate
in that context. It corresponds to the parameter ``U'' of Peskin
and Takeuchi \cite{peskin}.

We will focus on a special operator, that apart from breaking the
custodial symmetry, violates parity and charge conjugation while
conserving CP. The interest of this operator lies in the fact that
it is unique, and that violating parity, it can in principle
produce signatures that will set it apart from the other next-to-leading
terms in the effective Lagrangian. Furthermore, since the weak interactions
violate parity, there is no reason to expect this operator to have the
additional suppression factors usually associated with CP violation.
Observation of substantial effects in the custodial $SU(2)$ breaking
sector of the theory would have significant implications in our
understanding of electroweak symmetry breaking. In particular, models
would have to explain the smallness of $\Delta \rho$
in the absence of a custodial symmetry \cite{sik}.

In Section 2, we present the parity violating Lagrangian which is the
focus of our study and discuss the interactions which it generates.
In Sections 3 and 4, we turn our attention to potential future colliders.
Section 3 contains a particularly interesting result for asymmetries in
polarized $e^+ e^-$ production of $W$ boson pairs.  In Section 4
we estimate the size of the effect for $W Z$ production in an
$ e^- \gamma $ collider.
Rare $K$ and $B$ meson decays are discussed in Section 5.  We show that
the processes $B_s\ra \mu^+\mu^-$ and
$K^+\rightarrow \pi^+ \nu {\overline \nu}$
can be sensitive to the effects of the parity violating
operator in the effective Lagrangian.
Finally, Section 6 contains our conclusions.

\section{$P$ and $C$ violating but $CP$ conserving Lagrangian}

The effective Lagrangian with these properties is:
\beq
{\cal L} =  {\hat{\alpha}g v^2 \over \Lambda^2} \epsilon^{\alpha \beta
\mu \nu}\sp{\tau_3 \Sigma^\dagger D_\mu \Sigma}
\sp{W_{\alpha \beta} D_\nu \Sigma \Sigma^\dagger},
\label{leffpv}
\eeq
where $W_{\mu\nu}$ is the $SU(2)$ field strength tensor. In terms of
$W_\mu \equiv W^i_\mu \tau_i$, it is given by:\footnote{
Notice that there is a typo in Eq. 2.1 of Ref.~\cite{bdv}.}
\beq
W_{\mu\nu}={1 \over 2}\biggl(\partial_\mu W_\nu -
\partial_\nu W_\mu + {i \over 2}g[W_\mu, W_\nu]\biggr).
\eeq
A similar operator, with $B_{\alpha \beta}$ instead of $W_{\alpha \beta}$
would read:
\beqn
{\cal L} &=&  {\hat{\alpha}^\prime g^\prime
v^2 \over \Lambda^2} \epsilon^{\alpha \beta
\mu \nu}\sp{\tau_3 \Sigma^\dagger D_\mu \Sigma}
\sp{B_{\alpha \beta} D_\nu \Sigma^\dagger \Sigma} \nonumber \\
&=& {\hat{\alpha}^\prime g^\prime v^2 \over 2 \Lambda^2} \epsilon^{\alpha \beta
\mu \nu}B_{\alpha \beta} \sp{\tau_3 \Sigma^\dagger D_\mu \Sigma}
\sp{\tau_3 D_\nu \Sigma^\dagger \Sigma},
\eeqn
which is seen to vanish due to the antisymmetric nature of the epsilon tensor
after using $D_\nu \Sigma^\dagger \Sigma =- \Sigma^\dagger D_\nu\Sigma$.
Eq.~\ref{leffpv} is the only term in the effective
Lagrangian which violates parity and conserves $CP$ to ${\cal O}(E^4)$.

The operator of Eq.~\ref{leffpv} has been recently discussed
by Appelquist and Wu \cite{appel},
and the correspondence between our notation and theirs is
$\hat{\alpha}v^2/\Lambda^2 = \alpha_{11}$. The reason for the
additional factor that we introduce, is that this operator arises
at next-to-leading order (in the energy expansion), and is thus
suppressed by the scale of new physics.
In models where the operator is generated at
one-loop, as the one discussed in Ref.~\cite{appel}, the suppression
factor appears as $16 \pi^2$. This corresponds to the usual ``naive
dimensional analysis'' result $\Lambda \approx 4\pi v$.

It is instructive to consider the model of Ref.~\cite{appel}. In this
model, custodial $SU(2)$ is broken by the splitting between the masses
of new $SU(2)$ fermion doublets $m_U - m_D$, and the size of $\alpha_{11}$ is
constrained by $\Delta\rho$. Requiring the new physics to contribute
no more than a few percent to $\Delta\rho$, Ref.~\cite{appel} finds
$\alpha_{11} \leq 2 \times 10^{-4}$, which for $\Lambda = 1$~TeV,
corresponds to $\hat{\alpha} \leq 3 \times 10^{-3}$. In models where
$\Delta\rho$ is small as a consequence of an approximate custodial
symmetry, $\hat{\alpha}$ will have a natural size $\hat{\alpha}\approx
\Delta\rho$. This is consistent with the power counting analysis we sketched
in Ref.~\cite{bdv}. However, it is also possible, although not natural,
to have $\Delta\rho$ small without a custodial symmetry. In such models
$\hat{\alpha}$ would naturally be of order one.

In unitary gauge, the effects of the Lagrangian Eq.~\ref{leffpv},
are very simple. There is a three gauge boson interaction:
\beq
{\cal L}^{(3)} =-  {\hat{\alpha}g^3 v^2\over \Lambda^2 c_\theta}
\epsilon^{\alpha \beta \mu \nu}\biggl(W^-_\nu \partial_\alpha
W^+_\beta - W^+_\beta \partial_\alpha W^-_\nu\biggr)Z_\mu,
\label{agblt}
\eeq
which generates the $Z(q) \ra W^+(p^+) W^-(p^-)$ ``anomalous'' coupling
of Figure~1. In the notation of Ref.\cite{hagi} we have the
correspondence:
\beq
g_5^Z = \hat{\alpha}{g^2 \over c^2_\theta}{v^2\over \Lambda^2}.
\label{gfdef}
\eeq
There is also a four gauge boson interaction:
\beq
{\cal L}^{(4)}=i {2\hat{\alpha}g^4 v^2 s_\theta \over \Lambda^2 c_\theta}
\epsilon^{\alpha \beta \mu \nu}W^-_\alpha W^+_\beta
Z_\mu A_\nu.
\label{agblf}
\eeq
Some of the Feynman rules that can be derived from Eq.~\ref{leffpv},
are shown in Figure.~1. Our notation is $s_\theta = \sin{\theta_W}$,
$c_\theta = \cos{\theta_W}$.

Within the minimal standard model, the operator Eq.~\ref{leffpv} is
generated at one-loop by the splitting between top-quark and bottom-quark
masses. In the limit $m_t \gg m_W$, and setting $m_b=0$,
we find from the diagram in Figure~2:
\beq
\biggl({v^2 \over \Lambda^2}
\hat{\alpha}\biggr)_{\rm top} = {|V_{tb}|^2 \over 128 \pi^2}
\biggl(1-{8\over 3}s^2_\theta \biggr)\approx 3 \times 10^{-4}
\label{topc}
\eeq

Throughout our paper, we will express our results in terms of $g_5^Z$
adhering to convention. However, we wish to emphasize that the reader
should keep Eq.~\ref{gfdef}~in mind. This expression tells us
the natural size of $g_5^Z$, and its relation to the new physics
producing it. For example, if we assume that the new physics enters at
1~TeV, then $g_5^Z \sim {\cal O} (10^{-2})$ in theories in which there is
no custodial $SU(2)$ and $\Delta\rho$ is small accidentally. Similarly,
$g_5^Z \sim {\cal O}( 10^{-4})$ in theories that have an approximate
custodial $SU(2)$.

\section{Forward-backward asymmetry in $e^+_L e^-_R \ra W^+ W^-$}

In this section we study the effect of the parity violating operator
Eq.~\ref{leffpv} on the process $e^+ e^- \ra W^+ W^-$. This process
receives contributions from the diagrams of Figure~3. The $t$ channel
neutrino exchange diagram contributes only to $e^-_L e^+_R \ra W^- W^+$.
We will treat separately the two electron polarizations, because as we
will see, only the process with right-handed electrons generates an
observable proportional to $g_5^Z$.

We start by writing down the amplitudes generated by the lowest order
effective Lagrangian (Eq.~\ref{lagt} plus the kinetic energy terms for
the gauge fields), and by the parity violating Lagrangian Eq.~\ref{leffpv}.
For $e^-_R e^+_L$ we find:
\beqn
M(e^+_L e^-_R \ra W^+ W^-)= -{g^2 s^2_\theta \over s (s-m^2_Z)}
v_+^\alpha \epsilon^{*\mu}(p_3,\lambda^+)\epsilon^{*\nu}(p_4,\lambda^-)
\cdot \nonumber\\
\biggl[ m^2_Z\biggr(F_1^R g_{\mu\nu}(p_4 -p_3)_\alpha +
F_3^R(q_\mu g_{\alpha\nu} -q_\nu g_{\alpha\mu})\biggr)
+ i s F_5^R \epsilon_{\alpha\mu\nu\rho}(p_4-p_3)^\rho\biggr]
\eeqn
where $F_1^R=1$, $F_3^R =2$, and $F_5^R =g^Z_5$. For $e^-_L e^+_R$
we find:
\beqn
M(e^+_R e^-_L \ra W^+ W^-)= -
v_-^\alpha \epsilon^{*\mu}(p_3,\lambda^+)\epsilon^{*\nu}(p_4,\lambda^-)
\biggl\{{g^2 s^2_\theta \over s (s-m^2_Z)}
\cdot\nonumber\\
\biggl[m^2_Z \biggr( F_1^Lg_{\mu\nu}(p_4 -p_3)_\alpha +
F_3^L(q_\mu g_{\alpha\nu} -q_\nu g_{\alpha\mu})\biggr)
+ i s F_5^L \epsilon_{\alpha\mu\nu\rho}(p_4-p_3)^\rho\biggr]
\nonumber\\
+{g^2 \over 2t}(p_1-p_3)^\rho  \biggl( g_{\mu\rho}g_{\nu\alpha}
+g_{\nu\alpha}g_{\mu\rho}-g_{\alpha\rho}g_{\mu\nu}-i\epsilon_{\mu
\rho\nu\alpha}\biggr)\biggr\}
\eeqn
where now:
\beq
F_1^L = F_3^L/2 =
\biggl( 1 -{s\over m^2_Z}{1\over 2 s^2_\theta}\biggr) \;\;
F_5^L= g_5^Z \biggl(1-{1\over 2s^2_\theta}\biggr)
\eeq
and $s,t$ are the usual Mandelstam variables. We find it convenient to
use the vector equivalence technique \cite{eran}, in which the spinor
expression $\overline{v}_\pm(p_1)\gamma_\mu u_\pm(p_2)$ is replaced
with the equivalent vector $v_\pm =\sqrt{s} (0,1,\mp i,0)$ ($v_+$ being
a right-handed electron). This
allows computation of the amplitudes by explicitly replacing expressions
for all four-vectors in the $e^+ e^-$ center of
mass frame.

We then find that the differential cross-section contains a contribution
from the interference of the $g_5^Z$ term and the lowest order amplitude.
It also contains a contribution proportional to $|g_5^Z|^2$.
These contributions are present for both electron polarizations.
However, the cross-section for left-handed electrons is much larger
than the cross-section for right-handed electrons, and is not
very sensitive to the value of $g_5^Z$. This is why the studies of
unpolarized cross-sections in the literature have found the effect of
$g_5^Z$ to be less important than that of other (parity conserving)
anomalous couplings.

We will show that the cross-section with right-handed electrons is
much more sensitive to $g_5^Z$ than the unpolarized cross-section is.
However, deviations of the cross-section (polarized or not) from its
minimal standard model value can also be due to any of the parity
conserving anomalous couplings that we have ignored.

Of greater interest to us will be the fact that the parity violating
operator introduces a forward-backward asymmetry that is not present
in the minimal standard model for the case of right-handed electrons
(except, of course, for its one-loop contribution to $g_5^Z$ Eq.~\ref{topc}).
This forward-backward asymmetry is not
affected by the other anomalous couplings that we have ignored and it
is, therefore, the best place to search for $g_5^Z$.

The differential cross-section for right-handed electrons is given by:
\beqn
{d\sigma_{TT} \over d(\cos\theta)} \biggr|_{e^-_R}
&=& {\pi \alpha^2 \over  s} \beta^3
{m^4_Z \over (s-m^2_Z)^2}\sin^2\theta \nonumber \\
{d\sigma_{LL} \over d(\cos\theta)} \biggr|_{e^-_R}
&=& {\pi \alpha^2 \over 32 s}
{\beta^3 \over c^4_\theta} {s^2 \over (s-m^2_Z)^2}
(5+\beta^2)^2\sin^2\theta \nonumber \\
{d\sigma_{TL} \over d(\cos\theta)} \biggr|_{e^-_R}
&=& {\pi \alpha^2 \over  s}
{\beta^3 \over c^2_\theta}
{m^2_Z s\over (s-m^2_Z)^2}\biggl(1+\cos^2\theta + 2\beta{s\over m^2_Z}g^Z_5
\cos\theta\biggr)
\label{diffc}
\eeqn
where we use the notation $\beta^2=1-4m^2_W/s$.
We have summed over the different polarization states that contribute to
the cross sections with two transversely polarized $W$'s in the final state,
$\sigma_{TT}$, and with one transversely and one longitudinally polarized $W$'s
in the final state, $\sigma_{TL}$. Our result agrees with that of Ahn {\it et.
al.}\cite{ahn}.\footnote{Except for
what appears to be a typo in Eq.~2.10 of
Ref.~\cite{ahn} where we find that $A_3$ goes like $\beta$ and not like
$\beta^2$.} In terms of the notation of Ref.~\cite{ahn}, our result
contains only the tree-level standard model values of $F_1$ and $F_3$,
and we have only written terms that are linear in $g^Z_5$ (but our numerical
results also include the terms quadratic in $g_5^Z$). Other anomalous
couplings do not contribute to the forward backward asymmetry
in $e^-_R e^+_L \ra W^- W^+$ and they are not considered here.

As can be seen from Eq.~\ref{diffc}, there is a term in $\sigma_{TL}$ that is
linear in $\cos\theta$ (the scattering angle in the center of mass). This
term arises from the interference of $F_3$ and $F_5$ and gives rise to a
forward-backward asymmetry. Although there is a similar term in the
differential cross-section for $e^-_L e^+_R \ra W^+ W^-$, in that case one
also has a $t$-channel neutrino exchange diagram that gives rise to a very
large forward-backward asymmetry within the minimal standard model. Thus,
if we want to isolate the $g^Z_5$ term, it is very important to have
right-handed electrons. Since the cross-section for left-handed electrons
is several orders of magnitude larger than that for right-handed electrons,
it presents a formidable background. In Figure~4, we show the results
for the cross-section at $\sqrt{s}=200$~GeV, $\sqrt{s}=500$~GeV
and $\sqrt{s}=1$~TeV respectively. In these figures we assume that the
electron beam has a fraction $P_R$ of right-handed electrons and $(1-P_R)$ of
left-handed electrons. We can see that only the cross-section for
right-handed electrons is sensitive to the value of $g_5^Z$, and that
this sensitivity increases with increasing center of mass energy.

In Figure~5 we show the forward-backward asymmetry for $\sqrt{s}=200$~GeV,
$500$~GeV, and $1$~TeV. Again we find that the greatest sensitivity
to $g_5^Z$ occurs for right-handed electrons, and that this sensitivity
increases with increasing center of mass energy. However, in this case
we see that as long as one has a high degree of polarization, even the
lower energy machines could place a good bound on $g_5^Z$.

A detailed phenomenological study of this process would have to address
the issue of reconstruction of the scattering angle $\theta$ after the
$W$'s decay. It may also be possible to enhance the sensitivity to
$g_5^Z$ by using the fact that the forward-backward asymmetry is present
only in $\sigma_{TL}$.

\section{$e^- \gamma \ra \nu W^- Z$}

In this section we explore the possibility of observing the effects of
the parity violating operator Eq.~\ref{leffpv} via the anomalous
four-gauge-boson coupling that it generates. We thus turn our attention
to high energy vector-boson fusion experiments. Given the
form of the four vector-boson interaction, Eq.~\ref{agblf}, we look at
processes involving one photon and one $Z$. There are several possibilities,
for example $Z\gamma$ production in high energy $e^+ e^-$ or $pp$ colliders.
This process, however, suffers from large standard model backgrounds.
We will study instead an idealized situation where we can isolate the
effects of the new interaction
as much as possible from the backgrounds.
We consider a high energy $e^- \gamma$ collider where we can
cleanly identify the process $e^- \gamma \ra \nu W^- Z$, and where we
can also consider a polarized photon if need be. Some of the diagrams
that give rise to this process are shown in Figure~6.

The new interaction contributes both to the vector-boson fusion
diagrams, Figure 6a, and to the diagram that involves a three-gauge
boson vertex, Figure 6b. This interplay of three and four-gauge-boson
couplings from the same new operator makes the importance of a
gauge invariant formulation of the effective Lagrangian manifest.

As a first approximation, we will use the equivalence theorem to
replace the final state $W$ and $Z$ bosons by
their corresponding Goldstone bosons, $w$ and $z$.  We first
compute the process $W\gamma \rightarrow w z$.  The effective $W$
approximation is then used to fold the sub-process cross section with
the distribution of $W$'s in the electron \cite{effw}.

The leading order amplitude (generated from Eq.~\ref{lagt})
has been computed by us in Ref.~\cite{bdv}. To that contribution
we add the amplitude generated by Eq.~\ref{leffpv} to find (for $s>>m^2_W$):
\beqn
M(W^-(q_1)\gamma(q_2) \ra w^-(p_3) z(p_4))= \epsilon^\mu(q_1,\lambda^W)
\epsilon^\nu(q_2,\lambda^{\gamma})
g^2 s_\theta \nonumber \\
\biggl[ {2 \over us}\biggl({ut\over 2}g_{\mu\nu}
+up_{3\mu}q_{1\nu}+tq_{2\mu}p_{3\nu}+sp_{3\mu}p_{3\nu}\biggr)
-i {2 |g_5^Z| c_\theta^2 \over m_W^2}\epsilon_{\mu\nu\alpha\beta}q_2^\alpha
p^\beta_4 \biggr]
\label{wgppvbf}
\eeqn
The polarized cross sections, $\sigma (\lambda^W,\lambda^{\gamma})$,
are then:
\beqn
\sigma_{+-} &=\sigma_{-+}=&
{\pi\alpha^2 \over s^2_\theta}{1 \over 3s}
 \nonumber\\
\sigma_{++} &=\sigma_{--}=&
{\pi\alpha^2 \over s^2_\theta}{1 \over 3s}\biggl(
|g_5^Z|^2 c_\theta^4 {s^2\over m_W^4}\biggr) \nonumber\\
\sigma_{L+}&=\sigma_{L-}=&
{\pi\alpha^2 \over s^2_\theta}{1 \over 3s}\biggl(
|g_5^Z|^2 {c_\theta^4 \over 4}{s^3\over m_W^6}\biggr)
\label{swgpp}
\eeqn
We fold these cross-sections
with the luminosity for longitudinal and transverse $W$'s in an
electron to obtain the effective-W approximation result shown in
Figure~7. This figure indicates a potential sensitivity of this
process to values of $g_5^Z < 0.1$ which are within the interesting
range.

The subprocess cross-sections are identical
for the different photon polarizations if we sum over the $W$ polarization.
However, in the exact process $e^- \gamma \ra \nu w^- z$ the cross-section
depends on the photon polarization. Within the effective-$W$ approximation
this dependence is also present because the polarized cross-sections of
Eq.~\ref{swgpp} are weighted by
different factors: the distribution of $W$'s in the electron depends on
the $W$ polarization. This is also seen in Figure~7.

{}From Eq.~\ref{swgpp} we can see that the new term does not interfere with
the lowest order term: there is no contribution linear in $g_5^Z$. This
means that we can only construct observables sensitive to $g_5^Z$ that are
parity even and can thus be generated by other anomalous couplings.
Recall from Ref.~\cite{bdv}, that the
amplitude Eq.~\ref{wgppvbf} receives contributions from the next-to-leading
order operators $L_{9L}$, $L_{9R}$, and $L_{10}$; and that these contributions
do interfere with the leading amplitude. Nevertheless, it is possible that
the cross-section is more sensitive to the $|g_5^Z|^2$ term than to those
terms proportional to $L_{9L}$, $L_{9R}$ or $L_{10}$ in very high energy
machines. The reason for this is that the $|g_5^Z|^2$ term is
the only one that contributes to the amplitude where all three vector-bosons
are longitudinally polarized (this is the source of $\sigma_{L\pm}$ in
Eq.~\ref{swgpp}) and we expect these terms of ``enhanced electroweak
strength'' to dominate at high energies.

To construct an observable that can single out the $g_5^Z$
coupling we need a term in the differential cross-section linear in
$g_5^Z$. If we go beyond the effective-$W$ approximation, the new
term proportional to $g_5^Z$ will interfere with the lowest order
amplitude through the parity violating term in the fermionic structure
function \cite{han}. Going beyond the effective-$W$ approximation requires
the inclusion of the diagram in Figure~6b as well. The interference
term, not being present in the effective-W approximation, is thus
kinematically suppressed.

It appears that this process can potentially place significant constraints
on $g_5^Z$, but a detailed phenomenological study of the real process
$e^- \gamma \ra \nu W^- Z$ and its backgrounds is needed to draw any
conclusions.

\section{Rare $K$- and $B$-meson decays}

These rare decays receive contributions from the parity violating
effective Lagrangian Eq.~\ref{leffpv} at the one-loop level. One-loop
amplitudes with one vertex from the ${\cal O}(E^4)$ effective Lagrangian
are ${\cal O}(E^6)$. A complete study thus requires the next to next to
leading order counterterms, as well as two loop contributions from the
leading order effective Lagrangian. However, we will find that our
one-loop amplitudes are finite so we will be able to draw some
conclusions from our incomplete analysis. As a minimal consistency
check, we first look at the effects on the gauge boson
self-energies that could arise at the same order. This involves,
for example, the potential contributions to $\Delta\rho$ from
one-loop diagrams with one next-to-leading ($g_5^Z$) vertex.
However, one can easily see that there are no contributions to the
gauge boson self-energies linear in $g_5^Z$. This is evident, as there
are not enough independent four-vectors to saturate the indices of the
epsilon tensor. A contribution to the self-energies (and to $\Delta \rho$)
quadratic in $g_5^Z$ needs two next-to-leading vertices, and is therefore
one-order higher in perturbation theory (${\cal O}(E^8)$ in our notation).

As is well known, the effective operators responsible for rare meson
decays arise from box and penguin diagrams \cite{ina}. Since the lowest
order effective Lagrangian (complete with fermions), and the
new term Eq.~\ref{leffpv}, are separately gauge invariant, we are
free to treat the two terms independently. We argued that the
lowest order effective Lagrangian is just what remains when one
removes the Higgs-boson from the standard model by taking its mass
to infinity. However, it is easy to convince oneself that the
standard model operators responsible for rare $K$ and $B$ decays
do not depend on the Higgs-boson interactions. This is a consequence
of the usual approximation in which external quark masses and momenta are
set to zero. This means that, for example, Higgs-penguin diagrams
in which a Higgs-boson couples to $W$'s or to top-quarks vanish in the
limit of vanishing external quark masses and momenta. Since we will
work in this approximation, our lowest
order effective Lagrangian will simply reproduce the minimal standard
model results which are usually obtained in $R_\xi$ gauges.

As we said, the new term Eq.~\ref{leffpv} is separately gauge invariant,
so we may choose to perform
the calculations involving this term in any other gauge. The
simplest thing for us will be to perform them in unitary gauge. In this
gauge  Eq.~\ref{leffpv} enters only through
the anomalous $Z W^+ W^-$ coupling in the ``Z-penguin'' diagram
of Figure~8 at the one-loop level.

For a heavy top-quark, we can ignore the contributions of charm
and up-quarks in the intermediate state. The one-loop amplitude
that contributes to the rare decays is finite due to the GIM
cancelation as noted by He~\cite{he}. We obtain for the effective
one-loop vertex of Figure~8:
\beq
i\Gamma^\mu_{\rm PV}=-i{4 G_F \over \sqrt{2}}{\alpha \over 2 \pi s^2_\theta}
M^2_Z c_\theta V_{ti}V^*_{tj}\biggl(g\hat{\alpha}{v^2\over \Lambda^2}
\biggr) W(x_t)\overline{v}_i\gamma^\mu (1-\gamma_5)u_j
\label{penguin}
\eeq
where $x_t=m_t^2/m_W^2$ and we have defined
\beq
W(x_t)\equiv {3 \over 4}x_t\biggl({1\over 1- x_t}+{x_t \log x_t \over
(1-x_t)^2}\biggr)
\label{wxt}
\eeq
Our result agrees with that of Ref.~\cite{he}.
This contribution to the rare decays modifies the standard model results
for $K_L, B^0 \ra \ell^+ \ell^-$. In the notation of Ref.\cite{burasnew},
the full results (leading order plus new contribution) are obtained
by replacing:
\beqn
Y(x_t) &\ra & \hat{Y}(x_t)=
Y(x_t) + g_5^Z c^2_\theta W(x_t) \nonumber\\
Y(x_t) &=& {x_t \over 8}\biggl({x_t-4 \over x_t -1}+{3x_t \over (x_t-1)^2}
\log x_t\biggr)
\label{repl}
\eeqn

The case of $K_L \ra \mu^+ \mu^-$ was discussed by He \cite{he}. This
mode, however, has a large long distance contribution due to a two-photon
intermediate state that dominates the rate, and that is unaffected by
the new couplings. Although one
can compute reliably the absorptive part of the long distance component,
at present one cannot compute its dispersive part. It is therefore
not possible to place significant constraints on the short distance
component (and thus on $g_5^Z$) from the {\it measured} rate for this mode.
Thus, the constraint
obtained by He is purely theoretical, and it is equivalent to requiring
that the new contribution
be at most as large as the standard model short distance part. For
$m_t=150$~GeV, $Y(x_t)\approx W(x_t)$ so this implies:
\beq
g_5^Z \sim {\cal O}(1)
\label{heres}
\eeq
which is not a very stringent result if, as one expects,
$\Lambda \geq 1$~TeV.

A much better process to bound this contribution is $B_s \ra \mu^+ \mu^-$
because the rate is dominated by short distance physics, and is therefore
free of large theoretical uncertainties. This will allow us to obtain an
experimental bound on the anomalous coupling once this process is measured.
It will be a bound that can be improved by improving the accuracy of the
measurement. The rate for this process is
given by \footnote{Our standard model result agrees with that of
Ref.~\cite{sav,burasnew}.}:
\beq
\Gamma (B_s\ra \mu^+ \mu^-)={G_F^2 \over \pi}\biggl({\alpha \over 4 \pi
s^2_\theta}\biggr)^2F^2_B m^2_\mu m_B|V_{tb}V^*_{ts}|^2 \hat{Y}(x_t)^2
\label{bmumu}
\eeq
Numerically we use the Wolfenstein parameterization of the
CKM matrix with $A=.9$ and $\lambda=.22$. Our
normalization for $F_B$ is that
in which $f_\pi=132$~MeV, we use $F_B = 200$~MeV.
Although this process has a very small rate, it has a very clean
signature and should be seen in experiments at hadronic colliders with
vertex
detection. It is conceivable that a precision measurement of this rate
will exist in the future. In Figure~9 we have plotted the rate as a
function of $g_5^Z$. This figure confirms what one expects from
Eq.~\ref{repl}: a measurement of the rate to within factors of two
can only bound $g_5^Z$ to ${\cal O}(1)$. We can see from the figure that
the sensitivity to $g_5^Z$ increases with increasing top-quark mass.
It is also evident that a significant constraint
on $g_5^Z$ can only be placed by a precision measurement once the
top-quark mass and the CKM angles are known accurately.

The new vertex also contributes to the process $K^+ \ra \pi^+ \nu
\overline{\nu}$. This process is also dominated by short distance
physics so its precise measurement would allow us to place significant
constraints on $g_5^Z$ (or $\hat{\alpha}$). If we use the notation of Buras
{\it et. al.}\cite{buras} for the standard model result, we find the
full rate with the replacement:
\beqn
X(x_t) &\ra& \hat{X}(x_t)=X(x_t)
+ g_5^Z c^2_\theta W(x_t) \nonumber\\
X(x_t) &=& {x_t \over 8}\biggl({x_t+2 \over x_t -1}+{3x_t -6 \over (x_t-1)^2}
\log x_t\biggr)
\label{kpnnre}
\eeqn
in the contribution from a top-quark intermediate state, which becomes:
\beq
{B(K^+ \ra \pi^+ \nu \overline{\nu}) \over B(K^+ \ra \pi^0 e^+ \nu)}=
\biggl({\alpha \over \pi s^2_\theta}\biggr)^2 {|V_{td}V^*_{ts}|^2
\over |V_{us}|^2}\hat{X}(x_t)^2
\label{kpnnr}
\eeq
for each neutrino flavor. In Figure~10 we
have included the standard model charm-quark contribution with QCD
corrections as given in Ref.~\cite{buras} for typical values of all
unknown parameters. We see that this process will easily place bounds
of ${\cal O}(1)$ on $g_5^Z$, but that only a precision measurement combined
with detailed knowledge of the top-quark mass, CKM angles, and QCD corrections
could place significant constraints on $g_5^Z$.

As pointed out by He~\cite{he}, there is another
anomalous three-gauge-boson coupling, $g_1^Z-1$ in the notation
of Ref.~\cite{hagi}, that contributes to these processes at leading
order in $m_{ext}^2/m_W^2$. A deviation from the standard model rate
in these processes would, therefore, not be a definite signal for $g_5^Z$.

\section{Conclusions}

We have studied the possibility of observing the leading parity violating
operator in an effective Lagrangian description of the symmetry breaking
sector of the electroweak interactions. We have considered several observables
that are even under parity and that would not distinguish between the effect
of the parity violating interaction and a parity conserving one. We have also
studied one observable (the forward backward asymmetry in
$e^+_L e^-_R \ra W^+ W^-$) that would signal
exclusively the parity violating interaction.

The parity violating operator also breaks custodial
$SU(2)$ symmetry, and therefore its natural size depends on whether the
fundamental theory has a custodial symmetry or not. In theories with a
custodial symmetry (or an approximate one), we expect $g_5^Z$ to be ${\cal O}
(10^{-4})$ whereas without a custodial symmetry it could be
${\cal O}(10^{-2})$. The minimal standard model generates $g_5^Z$ at one-loop
at the $10^{-4}$ level.

The most promising place to look for a non-zero value of $g_5^Z$ is a
forward-backward asymmetry in polarized $e^-_R e^+_L$ collisions. The
sensitivity of this asymmetry to $g_5^Z$ is significantly reduced when
the polarization of the electron beam is not near 100\%. The asymmetry
is sensitive to $g_5^Z$ in machines with a center of mass energy as low
as 200~GeV, but a much better sensitivity is obtained at higher energies.
At higher energies, the total cross-section is also sensitive to $g_5^Z$
provided that there is a high degree of $e^-_R$ polarization.

We found that in addition to the usual anomalous three gauge boson vertex
associated with $g_5^Z$, gauge invariance requires the existence of a
four gauge boson vertex $\gamma Z W^+ W^-$ that is also proportional
to $g_5^Z$. We performed a preliminary study of the sensitivity of an
$e\gamma$ collider to $g_5^Z$ that makes use of this new coupling. We
find that at very high energies there is an increased sensitivity to
$g_5^Z$ because the new operator contains a coupling of the photon to
three longitudinal vector bosons not present in the minimal standard
model. The enhanced interactions of longitudinal vector bosons at
high energies are thus the origin of the potentially large sensitivity
of the process $e^- \gamma \ra \nu W^- Z$ to $g_5^Z$.

We find that the rare decays $B_s \ra \mu^+ \mu^-$ and $K^+ \ra \pi^+ \nu
\overline{\nu}$ can easily place bounds of ${\cal O}(1)$ on $g_5^Z$, but
that to improve this, one needs a precise measurement of the rate combined
with knowledge of all the standard model parameters.

A search for $g_5^Z$
in these observables would yield valuable information on the electroweak
symmetry breaking sector. In particular, an observation of a non-zero
$g_5^Z$ would be strong evidence against a custodial $SU(2)$ symmetry.

\noindent{\bf Acknowledgments}

We are grateful to Tao~Han, Xiao-Gang~He and Eran~Yehudai for
useful discussions.

\newpage

\noindent{\bf FIGURE CAPTIONS}

\begin{enumerate}

\item  Feynman rules from Eq.~\ref{leffpv}. We show the two vertices
that appear in unitary gauge, as well as some vertices involving would-be
Goldstone bosons that we use. The notation in the figure is
$s_\theta^2=\sin \theta_W^2=.23$, $c_\theta=\cos\theta_W$. Our convention
is that all momenta labelled $q$ enter into the vertex, and all labelled $p$
leave the vertex.

\item  One-loop contribution to $g_5^Z$ in the minimal standard model.

\item  Diagrams contributing to $e^+ e^- \ra W^+ W^-$. The full circle
in the first diagram represents the three gauge boson vertex both from
leading order and Eq.~\ref{leffpv}.

\item  Total cross-section for the process $e^+ e^- \ra W^+ W^-$ for
a) $\sqrt{s}=200$~GeV, b)$\sqrt{s} = 500$~GeV and c) $\sqrt{s}=1$~TeV.
The different curves from upper most to lowest correspond to a fraction
of right handed electrons in the beam of 0\%, 90\%, 95\%, 99\% and 100\%.

\item  Forward-backward asymmetry for the process $e^+ e^- \ra W^+ W^-$ for
a) $\sqrt{s}=200$~GeV, b)$\sqrt{s} = 500$~GeV and c) $\sqrt{s}=1$~TeV.
The different curves from upper most to lowest correspond to a fraction
of right handed electrons in the beam of 0\%, 90\%, 95\%, 99\% and 100\%.

\item  Types of diagrams contributing to $e^- \gamma \ra \nu W^- Z$.
a) Diagrams with the vector-boson fusion topology (including both
contact terms and $s$-and-$t$-channel gauge boson exchanges.
b) Diagram with a three gauge boson vertex that contributes to the
process $e^- \gamma \ra \nu w^- z$ beyond the effective-W approximation.

\item  $e^- \gamma \ra \nu w^- z$ cross-section in the effective-$W$
approximation.
We plot separately the results for each photon polarization
with $g_5^Z=0$ (lower curves) and with $g_5^Z = 0.1$ For $g_5^Z =0.1$
the upper curve corresponds to $\lambda^\gamma_+$ and the lower
curve to $\lambda^\gamma_-$. For $g_5^Z=0$ the upper curve corresponds
to $\lambda^\gamma_-$ and the lower curve to $\lambda^\gamma_+$.

\item  One-loop contribution from Eq.~\ref{leffpv} to the
$\overline{d}_i d_j Z$ effective vertex. The effective three gauge boson
vertex is represented by the full circle.

\item  Rate for $B_s \ra \mu^+ \mu^-$ as a function of $g_5^Z$. The dashed
curve corresponds to $m_t = 200$~GeV, the dotted curve to $m_t=150$~GeV
and the solid curve to $m_t=100$~GeV.

\item  $B(K^+ \ra \pi^+ \nu \overline{\nu})$ as a function of $g_5^Z$.
As an example we use $\rho=0$, $\eta=.4$, $V_{cb}=.041$,
$\Lambda_{QCD}=200~MeV$, and $m_c=1.4~GeV$
following Ref.~\cite{buras}. The dashed curve corresponds to
$m_t=200$~GeV, the dotted curve to $m_t=150$~GeV and the solid curve to
$m_t=100$~GeV.

\end{enumerate}

\end{document}